# A handy systematic method for data hazards detection in an instruction set of a pipelined microprocessor

Ahmed M. Mahran

Computer and Systems Engineering Department, Faculty of Engineering,
Alexandria University, Alexandria 21544, Egypt
Email: ahmahran@gmail.com

***Abstract*** *– It is intended in this document to introduce a handy systematic method for enumerating all possible data dependency cases that could occur between any two instructions that might happen to be processed at the same time at different stages of the pipeline. Given instructions of the instruction set, specific information about operands of each instruction and when an instruction reads or writes data, the method could be used to enumerate all possible data hazard cases and to determine whether forwarding or stalling is suitable for resolving each case.*

## I. INTRODUCTION

Computer architecture students who study the pipelined microprocessor (like MIPS architecture) usually end with implementing a microprocessor that supports a subset of the instruction set. The most cumbersome and tricky part of the implementation process is the part of enumerating all data hazards that are possible to occur among the selected instructions while being processed at the same time at different stages of the pipeline. Usually, students investigate simple code snippets trying as much mixes of instructions as possible to extract out different data hazard cases. The more test cases are provided, the more reliable the design becomes but the process becomes more complicated and error prone. There is not a systematic routine that could be followed to simplify the process. However, there are techniques used to automate the design process given some kind of description of the required microprocessor but those techniques are not suitable for pedagogical purposes. Learners need to get their hands dirty and carry out steps by hand to develop the sense. Hence, there is a need for a handy systematic method to organize the process.

It is intended in this document to introduce a handy systematic method for enumerating all possible data dependency cases that could occur between any two instructions that might happen to be processed at the same time at different stages of the pipeline. Given instructions of the instruction set, specific information about operands of each instruction and when an instruction reads or writes data, the method could be used to enumerate all possible data hazard cases and to determine whether forwarding or stalling is suitable for resolving each case.

## II. DATA HAZARDS

There are three types of hazards that need to be resolved in a pipelined microprocessor implementation. They are data hazards, control hazards and structural hazards. This document is concerned with the detection of data hazards only. Depending on the order of read and write accesses of instructions, data hazards are classified into three types:

- Read After Write (RAW)
- Write After Read (WAR)
- Write After Write (WAW)

Some information about each instruction should be known in order to be able to detect each data hazard. The following subsections investigate this issue.

**Read After Write (RAW) hazard**

A read after write (RAW) hazard occurs when a newer instruction (i) depends on the result of an earlier instruction (j) such that the result of the earlier instruction (j) hasn't been written to its final destination yet and still the new instruction (i) needs to read the result from its agreed final destination.

Let's call the earlier instruction (j) the producer instruction and the newer instruction (i) the consumer instruction. So, a RAW hazard happens when a consumer instruction reaches a stage at which it requires to consume data either produced or will be produced by an earlier producer instruction such that the data is not yet saved to its expected final destination. So, it's important to know those stages for each consumer instruction.

Generally, it is important to know when data is available from the producer and when it is needed by the consumer so that one can know at which stages to look for hazard problems. Also, knowing when data is available from the producer helps in determining when it is possible to start the solution for a hazard problem. Besides, knowing when data is needed by the consumer helps in determining when it is critical to have the solution applied.

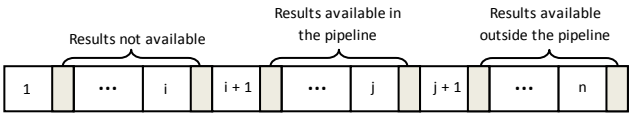

**Figure 1 - Different periods of a producer instruction in a pipeline**

Figure 1 shows the n stages of a pipeline being grouped into three different intervals. The start and the end of each interval differ from one producer instruction to another. The first interval from stage 1 to stage i is the interval when the results of the producer instruction are not yet available. The second interval starts with stage i + 1 when the results become available and ends at stage j that is the last stage for results to be available at the pipeline. The third interval starts at stage j + 1 and ends at the last stage n during which the results are not available inside the pipeline but could be retrieved from outside the pipeline's stage registers at their expected final destination.

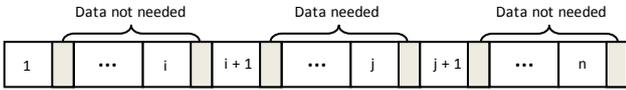

**Figure 2 - Different periods of a consumer instruction in a pipeline**

For a consumer instruction (as shown in Figure 2) there are three intervals. It is considered too early to provide the data for the consumer instruction at the stages of the first interval. At the second interval, source data is needed by the consumer instruction and could be provided to the consumer instruction at any stage of this interval. At the third interval, the data becomes not needed any more.

The most important intervals for the proposed method are the result's availability interval for each producer instruction and the data's need interval for each consumer instruction.

**Write After Read (WAR) hazard**

A write after read (WAR) hazard occurs when a newer instruction (i) wants to overwrite a destination before an earlier instruction (j) reads it such that the earlier instruction gets the wrong data.

Let's call the newer instruction (i) the writer instruction and the earlier instruction (j) the reader instruction. The writer instruction should overwrite its destination after the reader instruction finishes reading it. So, it is important to know the stage when the writer instruction writes data and stage when the reader instruction reads data.

**Write After Write (WAW) hazard**

A write after write (WAW) hazard occurs when a newer instruction (i) wants to write to a destination before an earlier instruction (j) writes it such that the destination is wrongly updated.

Both instructions are writer instructions so let's call the newer instruction (i) the second writer instruction and the earlier instruction (j) the first writer instruction. The second writer instruction should update its destination after the first instruction finishes updating it. So, it is important to know the stages when both writer instructions write data.

### III. THE METHOD

The method is targeting any instruction set consisting of instructions having $P$ data sources and $Q$ data destinations (i.e. $P + Q$ operands) where $P$ and $Q$ are any positive integers greater than or equal to zero and could differ from one instruction to another.

In order to understand how the method works, the following simple definitions are introduced first.

**Execution sequence:** an instruction's execution sequence/path is an ordered sequence of pipeline stages that the instruction passes through while being processed in the pipeline. Consider an instruction that enters the pipeline at stage 1 passing though all stages till it leaves the pipeline at stage $k$. The execution sequence is then $\langle 1, 2, \ldots, k-1, k \rangle$.

**Coupled/Paired execution sequence:** it is an ordered sequence of ordered pairs of pipeline stages of two instructions being processed concurrently in the pipeline. Consider two instructions, *inst1* which has the execution sequence $\langle 1,2,3,4,5 \rangle$ and *inst2* which has the execution sequence $\langle 1,2,3 \rangle$. The coupled execution sequence when *inst2* enters the pipeline the next cycle after *inst1* enters (assuming no hazards) is then $\langle (1,2), (2,3), (3,4) \rangle$. The first pair $(1,2)$ means that *inst2* is at stage 1 while *inst1* is at stage 2. Then, at the second cycle, *inst2* advances to stage 2 while *inst1* advances to stage 3 and this is reflected by the pair $(2,3)$. After one more cycle, *inst2* is advanced to its final stage, stage 3, while *inst1* is advanced to stage 4. It should be noticed that stages 1 and 5 of *inst1* are not included in the coupled sequence that's because *inst2* is not in the pipeline while *inst1* is at stages 1 and 5 hence both stages have no associate stages of *inst2* to be coupled with hence the coupled sequence begins with the pair $(1,2)$ and not $(-,1)$ and ends with the pair $(3,4)$ and not with $(-,5)$.

In general, the method tries all possible pairs of instructions inspecting all possible coupled execution sequences for all possible data hazard cases. Figure 3 shows a graphical representation of all possible coupled execution sequences between *inst1* and *inst2* when *inst2* enters the pipeline after *inst1*. Each arrow represents one possible sequence while each small square represents an ordered pair of stages belonging to the sequence corresponding to the arrow drawn on it. The longest arrow rep-resents the sequence ⟨(1,2),(2,3),(3,4)⟩ which is mentioned earlier.

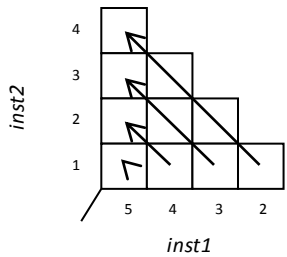

**Figure 3 - A graphical representation of all possible coupled execution sequences of inst1 and inst2 as inst2 is the recent instruction**

### Step I

In step I, the following data should be prepared for each instruction of the instruction set.

| inst. | Operand | R/W | First needed/available | Last needed/available |
|---|---|---|---|---|
| opcode | $s_1$ | | | |
| | . | | | |
| | . | | | |
| | $s_P$ | | | |
| | $d_1$ | | | |
| | . | | | |
| | . | | | |
| | $d_Q$ | | | |

**Table 1 - data required to be prepared at step I**

The first column (named "inst.") determines the instruction's opcode. The second column (named "Operand") determines the operands of the instruction and whether each is a source or a destination operand. The third column (named "R/W") determines the stage at which the corresponding operand is being read or written if it is a source or a destination operand respectively. The fourth column (named "First needed/available") determines the first stage at which the value of the corresponding operand is either needed (if it is a source operand) or available at the pipeline (if it is a destination operand). The fifth column (named "Last needed/available") determines the last stage at which the value of the corresponding operand is either needed (if it is a source operand) or available at the pipeline (if it is a destination operand).

At the end of this step, data of all instructions are grouped in one table.

### Step II

This step is concerned with reducing the table formed at step I. The reduction is performed by grouping similar instructions into one group and similar source (or destination) operands into one group of sources (or destinations) for each instruction. Either type of similarity is determined by the following two rules.

- For one instruction, two source (destination) operands are equivalent if and only if:
  o They are read (written) at the same stage.
  o They are first needed (available) at the same stage.
  o They are last needed (available) at the same stage.

- Two instructions are equivalent if and only if:
  o The source (destination) operands of a one are equivalent in a one to one correspondence manner to the source (destination) operands of the other.

### Step III

RAW hazards are determined at this step but first the table from step II is reduced again by applying the following equivalence rules:

- For an instruction, two source (destination) operands are equivalent if and only if:
  o They are first needed (available) at the same stage.
  o They are last needed (available) at the same stage.

- Two instructions are equivalent if and only if:
  o The source (destination) operands of a one are equivalent in a one to one correspondence manner to the source (destination) operands of the other.

After that, for each possible pair of a consumer and a producer instruction, all possible coupled execution sequences are inspected for RAW hazards different times. Each time, a different source operand of the consumer instruction is considered equal to a different destination operand of the producer instruction during a coupled execution. If a RAW hazard is detected, it could be determined either a forward or a stall is required to resolve the problem when the pipeline is in the state of having the corresponding configurations.

Consider the following information about *inst1* and *inst2*:

| inst. | Operand | R/W | First needed/available | Last needed/available |
|---|---|---|---|---|
| inst1 | $s_1$ | - | - | - |
| | $s_2$ | - | - | - |
| | $d_1$ | 4 | 3 | 4 |
| | $d_2$ | 5 | 5 | 5 |
| inst2 | $s_1$ | 1 | 1 | 1 |
| | $s_2$ | 2 | 1 | 2 |
| | $d_1$ | - | - | - |
| | $d_2$ | - | - | - |

**Table 2 - Prepared data for sample instructions from a hypothetical instruction set**

Let's consider *inst2* the consumer instruction which enters the pipeline after *inst1* that is the producer instruction. During any coupled execution of both instructions, there are four possibilities of a source operand of *inst2* being equal to a destination operand of *inst1*. These cases are:

1) $inst1.d_1 = inst2.s_1$
2) $inst1.d_1 = inst2.s_2$
3) $inst1.d_2 = inst2.s_1$
4) $inst1.d_2 = inst2.s_2$

For each case, all possible coupled execution sequences are inspected for the possibility of a RAW hazard. A RAW hazard occurs when a consumer instruction cannot advance to the next pipeline stage at the next cycle because an earlier producer instruction hasn't put the required data at its designated final destination yet. That means a RAW hazard occurs when the consumer instruction is at the stage when the source operand is last needed and data is not yet available at the destination operand of the producer instruction i.e. the producer instruction is at one of the stages from the stage at which it first enters the pipeline to the stage when data is last available in the pipeline registers i.e. a stage from either the first or second interval shown in Figure 1. If the producer instruction has already produced the data i.e. it is at one of the stages in the data availability interval (the second interval in Figure 1), then a forward is required to bypass the required data to the consumer instruction. However, if the producer instruction hasn't produced data yet i.e. it is at a stage from the first interval (which is illustrated in Figure 1), then the consumer instruction should be stalled until data is available -i.e. until the producer instruction enters the second interval- then data forwarding is applied. Otherwise data would be available at its designated destination when the producer instruction enters the third interval and no action is needed then as there is no hazard in this case.

Continuing with the example, let's inspect each of the previously mentioned four cases. Figure 4 is used as a graphical aid.

1) $inst1.d_1 = inst2.s_1$

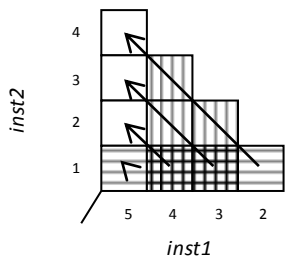

**Figure 4 – Inspection of coupled sequences for RAW hazards when $inst1.d_1 = inst2.s_1$**

As shown in Figure 4, there are four possible coupled execution sequences. As mentioned before, A RAW hazard occurs when the consumer instruction (*inst2*) is at the stage at the end of the second interval and the producer instruction (*inst1*) is at a stage from either the first or the second interval. If the producer instruction is at the first interval, then a stall is required. If it is at the second interval, then a forward is required. Applying these rules, the first row from the bottom in Figure 4 corresponds to steps in the execution sequences when *inst2* (the consumer) is at the end of the second interval. Whereas column marked 2 corresponds to steps in the execution sequences when *inst1* is at its first interval and columns marked 3 and 4 correspond to the second interval. In the light of the foregoing, RAW hazards occur at squares in the intersections between the row marked 1 and columns marked 2, 3 and 4. The result of the first intersection is the first step in the first (longest) execution sequence that corresponds to the pair (1,2) while the second and third intersections result in the pairs (1,3) and (1,4) respectively. For the pair (1,2), *inst1* would produce data in the next stage, hence *inst2* should be stalled for one cycle. For the pair (1,3), *inst1* already produced data, hence data is forwarded from stage 3 to stage 1. Also, data is forwarded from stage 4 to stage 1 for the pair (1,4).

2) $inst1.d_1 = inst2.s_2$

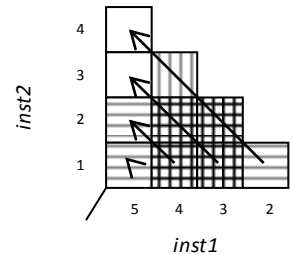

**Figure 5 - Inspection of coupled sequences for RAW hazards when $inst1.d_1 = inst2.s_2$**

In Figure 5, the end of the second interval of *inst2* intersects with the second interval of *inst1* and has no intersection with the first interval. Hence, there are RAW hazards to be resolved by forwarding only and they are at steps (2,3) and (2,4) such that data is forwarded from stage 3 and stage 4 to stage 2, respectively.

3) $inst1.d_2 = inst2.s_1$

In Figure 6, the end of the second interval of inst2 intersects with the first interval of inst1 at the pairs (1,2), (1,3) and (1,4) hence inst2 should be stalled for 3 cycles, 2 cycles and 1 cycle, respectively. Also, the end of the second interval of inst2 intersects with the second interval of inst1 at the pair (1,5) hence a forward is required from stage 5 to stage 1.

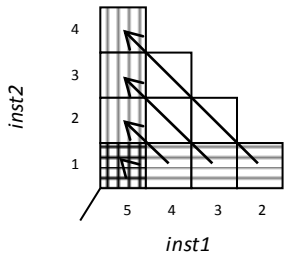

**Figure 6 - Inspection of coupled sequences for RAW hazards when $inst1.d_2 = inst2.s_1$**

4) $inst1.d_2 = inst2.s_2$

In Figure 7, the end of the second interval of inst2 intersects with the first interval of inst1 at the pairs (2,3) and (2,4) hence inst2 should be stalled for 2 and 1 cycles, respectively. Also, the end of the second interval of inst2 intersects with the second interval of inst1 at the pair (2,5) hence a forward is required from stage 5 to stage 2.

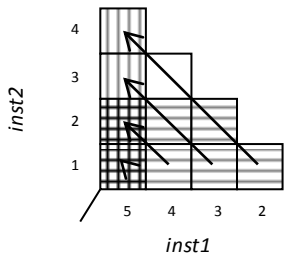

**Figure 7 - Inspection of coupled sequences for RAW hazards when $inst1.d_2 = inst2.s_2$**

As we are interested in the end of the second interval of the consumer instruction only and not in the whole interval, the fourth column of Table 1 (named "First needed/available") could be ignored for consumer instructions only so that it is not considered by the equivalence rules mentioned before. As a result also, Figure 4 and Figure 5 could be condensed in one figure as shown in Figure 8. The same goes for Figure 6 and Figure 7 as well.

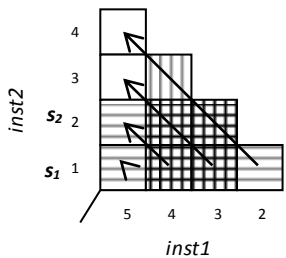

**Figure 8 - Inspection of coupled sequences for RAW hazards when $inst1.d_1 = inst2.s_1$ and $inst1.d_1 = inst2.s_2$**

**Step IV**

WAR hazards are determined at this step but first the table from step II is reduced again by applying the following equivalence rules:

- For an instruction, two source (destination) operands are equivalent if and only if:
  o They are read (written) at the same stage.

- Two instructions are equivalent if and only if:
  o The source (destination) operands of a one are equivalent in a one to one correspondence manner to the source (destination) operands of the other.

After that, for each possible pair of a reader and a writer instruction, all possible coupled execution sequences are inspected for WAR hazards different times. Each time, a different source operand of the reader instruction is considered equal to a different destination operand of the writer instruction during a coupled execution.

| inst. | Operand | R/W | First needed/available | Last needed/available |
|---|---|---|---|---|
| inst1 | $s_1$ | - | - | - |
| inst1 | $s_2$ | - | - | - |
| inst1 | $d_1$ | 1 | - | - |
| inst1 | $d_2$ | 2 | - | - |
| inst2 | $s_1$ | 4 | - | - |
| inst2 | $s_2$ | 5 | - | - |
| inst2 | $d_1$ | - | - | - |
| inst2 | $d_2$ | - | - | - |

**Table 3 - Prepared data for sample instructions from another hypothetical instruction set**

For example, Table 4 shows information of another instruction set. Consider *inst2* the reader instruction which enters the pipeline after *inst1* that is the writer instruction. During any coupled execution of both instructions, there are a number of possibilities of a source operand of *inst2* being equal to a destination operand of *inst1*. These cases are:

1) $inst1.d_1 = inst2.s_1$
2) $inst1.d_1 = inst2.s_2$
3) $inst1.d_2 = inst2.s_1$
4) $inst1.d_2 = inst2.s_2$

For each case, all possible coupled execution sequences are inspected for the possibility of a WAR hazard. A WAR hazard occurs when a writer instruction must not write a destination operand before an earlier reader instruction reads it first. That means a WAR hazard occurs when the writer instruction is at the last stage before the stage when data is written and the reader instruction is at any stage before the stage when data is read.

Continuing with the example, Figure 9 is used as a graphical aid and Table 4 gives a summary of all WAR hazards.

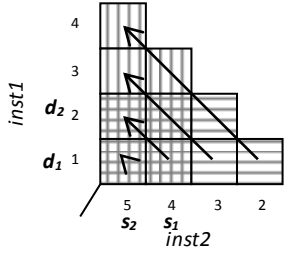

**Figure 9 - Inspection of coupled sequences for WAR hazards**

| Case | Hazard | Stalled inst. | # of stall cycles |
|---|---|---|---|
| inst1.d1 = inst2.s1 | (1,2) | inst1 | 3 |
|  | (1,3) | inst1 | 2 |
|  | (1,4) | inst1 | 1 |
| inst1.d2 = inst2.s1 | (2,3) | inst1 | 2 |
|  | (2,4) | inst1 | 1 |
| inst1.d1 = inst2.s2 | (1,2) | inst1 | 4 |
|  | (1,3) | inst1 | 3 |
|  | (1,4) | inst1 | 2 |
|  | (1,5) | inst1 | 1 |
| inst1.d2 = inst2.s2 | (2,3) | inst1 | 3 |
|  | (2,4) | inst1 | 2 |
|  | (2,5) | inst1 | 1 |

**Table 4 – A summary of detected WAR hazards**

### Step V

This step is very similar to the previous step except that WAW hazards are to be determined. The same procedures should be followed taking into consideration that two writer instructions are paired instead of a reader and a writer instructions.

For the example of Table 3, *inst1* is paired with itself. Figure 10 is used as a graphical aid and Table 5 gives a summary of all WAW hazards. Columns correspond to *inst1* as a first writer and rows correspond to *inst1* also but as a second writer.

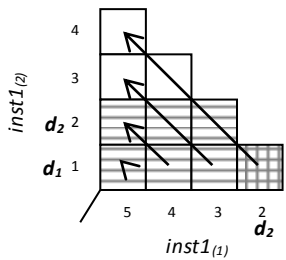

**Figure 10 - Inspection of coupled sequences for WAW hazards**

| Case | Hazard | Stalled inst. | # of stall cycles |
|---|---|---|---|
| $inst1_{(1)}.d1 = inst1_{(2)}.d1$ | - | - | - |
| $inst1_{(1)}.d1 = inst1_{(2)}.d2$ | - | - | - |
| $inst1_{(1)}.d2 = inst1_{(2)}.d1$ | (1,2) | $inst1_{(2)}$ | 1 |
| $inst1_{(1)}.d2 = inst1_{(2)}.d2$ | - | - | - |

**Table 5 - A summary of detected WAW hazards**

## IV. FORMALIZATION

**Step I** – For an instruction set $IS$ to be implemented as a pipelined microprocessor with $n$ pipeline stages, data in Table 6 should be prepared for each instruction $I_i \in IS$.

| inst. | Operand | R/W | First needed/available | Last needed/available |
|---|---|---|---|---|
| $I_i$ | $s_p^i$ | $r_p^i$ | - | $ln_p^i$ |
|  | $d_q^i$ | $w_q^i$ | $fa_q^i$ | $la_q^i$ |

**Table 6 - Prepared data for instructions of the instruction set IS**

Such that, $s_p^i \in S_i$ and $d_q^i \in D_i$ where $S_i$ and $D_i$ are the sets of source and destination operands of instruction $I_i$, respectively and $p = 1, 2, \ldots, |S_i|$ and $q = 1, 2, \ldots, |D_i|$. Also, $r_p^i, ln_p^i, w_q^i, fa_q^i, la_q^i = 1, 2, \ldots, n$ and $la_q^i \geq fa_q^i$.

**Step II** – The prepared table is reduced by applying the following equivalence rules to get the reduced instruction set $IS^1$. First, $S_i$'s and $D_i$'s are reduced to $S_i^1$'s and $D_i^1$'s, respectively:

$$\forall I_i \in IS, \forall s_k^i, s_l^i \in S_i; s_k^i \equiv s_l^i \leftrightarrow (r_k^i = r_l^i \land ln_k^i = ln_l^i)$$

$$\forall I_i \in IS, \forall d_k^i, d_l^i \in D_i; d_k^i \equiv d_l^i \leftrightarrow (w_k^i = w_l^i \land fa_k^i = fa_l^i \land la_k^i = la_l^i)$$

Then, $IS$ is reduced to $IS^1$:

$$\forall I_i, I_j \in IS; I_i \equiv I_j \leftrightarrow \left( (\forall s_k^i \in S_i^1 \exists s_l^j \in S_j^1; s_k^i \equiv s_l^j \land |S_i^1| = |S_j^1|) \\ \land (\forall d_k^i \in D_i^1 \exists d_l^j \in D_j^1; d_k^i \equiv d_l^j \land |D_i^1| = |D_j^1|) \right)$$

**Step III** – The reduced table from step II is reduced again by applying the following equivalence rules to get the reduced instruction set $IS^2$. First, $S_i^1$'s and $D_i^1$'s are reduced to $S_i^2$'s and $D_i^2$'s, respectively:

$$\forall I_i \in IS^1, \forall s_k^i, s_l^i \in S_i^1; s_k^i \equiv s_l^i \leftrightarrow (ln_k^i = ln_l^i)$$

$$\forall I_i \in IS^1, \forall d_k^i, d_l^i \in D_i^1; d_k^i \equiv d_l^i \leftrightarrow (fa_k^i = fa_l^i \land la_k^i = la_l^i)$$

Then, $IS^1$ is reduced to $IS^2$:

$$\forall I_i, I_j \in IS^1; I_i \equiv I_j \leftrightarrow \left( (\forall s_k^i \in S_i^2 \exists s_l^j \in S_j^2; s_k^i \equiv s_l^j \land |S_i^2| = |S_j^2|) \\ \land (\forall d_k^i \in D_i^2 \exists d_l^j \in D_j^2; d_k^i \equiv d_l^j \land |D_i^2| = |D_j^2|) \right)$$

After that, each RAW hazard is determined as a pair of stages corresponding to the step in an execution sequence when the hazard occurs.

$\forall I_i, I_j \in IS^2, \forall d_q^i \in D_i^2 \forall s_p^j \in S_j^2$ such that $I_i$ is a producer instruction and $I_j$ is a consumer instruction that enters the pipeline after $I_i$, Table 7 gives rules for RAW hazards enumeration.

| Hazard | Soln. | Apply soln. at |
|---|---|---|
| $(ln_p^j, k)$ $\forall k \in [fa_q^i, la_q^i] \wedge k > ln_p^j$ | forward from stage k to stage $ln_p^j$ | $(ln_p^j, k)$ |
| $(ln_p^j, k)$ $\forall k \in [1, fa_q^i[ \wedge k > ln_p^j$ | stall # stalls = $fa_q^i - k$ | $(1, k - ln_p^j + 1)$ |

Table 7 – Rules for RAW hazards enumeration

**Step IV** – The reduced table from step II is reduced again by applying the following equivalence rules to get the reduced instruction set $IS^3$. First, $S_i^{1}$'s and $D_i^{1}$'s are reduced to $S_i^{3}$'s and $D_i^{3}$'s, respectively:

$$\forall I_i \in IS^1, \forall s_k^i, s_l^i \in S_i^1; s_k^i \equiv s_l^i \leftrightarrow (r_k^i = r_l^i)$$

$$\forall I_i \in IS^1, \forall d_k^i, d_l^i \in D_i^1; d_k^i \equiv d_l^i \leftrightarrow (d_k^i = d_l^i)$$

Then, $IS^1$ is reduced to $IS^3$:

$$\forall I_i, I_j \in IS^1; I_i \equiv I_j \leftrightarrow \left((\forall s_k^i \in S_i^3 \exists s_l^j \in S_j^3; s_k^i \equiv s_l^j \wedge |S_i^3| = |S_j^3|) \right.$$
$$\left. \wedge (\forall d_k^i \in D_i^3 \exists d_l^j \in D_j^3; d_k^i \equiv d_l^j \wedge |D_i^3| = |D_j^3|)\right)$$

After that, each WAR hazard is determined as a pair of stages corresponding to the step in an execution sequence when the hazard occurs.

$\forall I_i, I_j \in IS^3, \forall s_p^i \in S_i^3 \forall d_q^j \in D_j^3$ such that $I_i$ is a reader instruction and $I_j$ is a writer instruction that enters the pipeline after $I_i$, Table 8 gives rules for WAR hazards enumeration.

| Hazard | Soln. | Apply soln. at |
|---|---|---|
| $(w_q^j, k)$ $\forall k \in [1, r_p^i[ \wedge k > w_q^j$ | stall # stalls = $r_p^i - k + 1$ | $(1, k - w_q^j + 1)$ |

Table 8 – Rules for WAR hazards enumeration

**Step V** – The reduced instruction set $IS^3$ from step IV is used here. Each WAW hazard is determined as a pair of stages corresponding to the step in an execution sequence when the hazard occurs.

$\forall I_i, I_j \in IS^3, \forall d_p^i \in D_i^3 \forall d_q^j \in D_j^3$ such that $I_i$ is a first writer instruction and $I_j$ is a second writer instruction that enters the pipeline after $I_i$, Table 9 gives rules for WAW hazards enumeration.

| Hazard | Soln. | Apply soln. at |
|---|---|---|
| $(w_q^j, k)$ $\forall k \in [1, w_q^i] \wedge k > w_q^j$ | stall # stalls = $w_q^i - k + 1$ | $(1, k - w_q^j + 1)$ |

Table 9 – Rules for WAW hazards enumeration

## V. CONCLUSION AND FUTURE WORK

It is introduced in this document a systematic method to enumerate data hazards and their solutions for an instruction set that would be implemented as a pipelined microprocessor. The method is simple and could be carried by hand which make it a good candidate for pedagogical purposes. Students' work would be more organized and less painful.

It's not investigated here how this systematic method could be exploited in the automatic design of the forwarding and the stalling units of the pipeline which could be considered as a future work.

## VI. ACKNOWLEDGMENT